\documentclass[
preprint,
prd,
showpacs,
amsmath,
amssymb,
floatfix,
nofootinbib,
]{revtex4}
\usepackage[dvips]{graphicx}

\begin{document}

\title{Black hole formation in the head-on collision \\ of ultrarelativistic charges}

\author{Hirotaka Yoshino$^{(1,2)}$}
\author{Robert B. Mann$^{(3)}$}


\affiliation{$^{(1)}$ Department of Physics, Tokyo Institute of Technology, Tokyo 152-8551, Japan}

\affiliation{$^{(2)}$ Graduate School of Science and Engineering, Waseda University, Shinjuku-ku, Tokyo 169-8555, Japan}

\affiliation{$^{(3)}$ Department of Physics, University of Waterloo, 200 University Avenue West, Waterloo, Ontario N2L 3G1, Canada}

\date{May 26, 2006}

\begin{abstract}
We study  black hole formation in the head-on collision
of ultrarelativistic charges. 
The metric of charged particles is obtained by boosting the
Reissner-Nordstr\"om spacetime to the speed of light.
Using the slice at the instant of collision, we study formation of the apparent horizon (AH)
and derive a condition indicating that a critical value of the electric charge is necessary
for formation to take place.
Evaluating this condition for characteristic values at the LHC,
we find that the presence of charge
decreases the black hole production rate in accelerators.
We comment on possible limitations of our approach.
\end{abstract}

\maketitle


\section{Introduction}

Black holes are expected to form in the collisions of elementary
particles with energies above the Planck scale. It has been pointed out that
the Planck energy could be $O$(TeV) if our space is a 3-brane situated in a large extra dimensional
space  and  gauge particles and interactions are confined on the brane~\cite{ADD98}.
If such TeV gravity scenarios are realized, we would be able to directly observe
black hole phenomena in planned accelerators such as the Large
Hadron Collider (LHC) at CERN.  The general scenario is expected to be
as follows \cite{BF99, DL01, GT02}. First  the horizon forms (the
black hole production phase), after which
the black hole is expected to go to a stationary Kerr black hole
by radiating gravitational waves (the balding phase).
Then it will evaporate via Hawking radiation (the evaporation phase).
 In the context of this three-phase scenario, the main problems of interest are
the black hole production rate (see \cite{DL01,GT02} and also
\cite{EG02, YN03, GR04,YR05,CBC05}),
the determination of the mass and the angular momentum
of the  Kerr black hole (see \cite{DP92, GW-emission, CBC05} for related issues),
and the prediction of Hawking radiation \cite{greybody}.  The 
interactions between the produced black holes and the brane are
discussed in \cite{brane}. See also \cite{reviews} for reviews.

In this paper, we consider an issue related to the
black hole production rate. The production rate at the LHC
was first calculated in \cite{DL01,GT02}.
In  proton collisions at the LHC, their constituent partons will collide and
form a black hole. The cross section in the parton collision
was assumed to be $\sigma_{\rm BH}\simeq \pi [r_h(2p)]^2$,
where $r_h(2p)$ is the gravitational radius corresponding to the
system energy $2p$.  Integrating this by multiplying the parton distribution
functions, the total cross section is derived  upon summing all possible parton pairs.
Black hole production rate is about 1Hz under this assumption.

\begin{figure}[bt]
\centering
{
\includegraphics[width=0.25\textwidth]{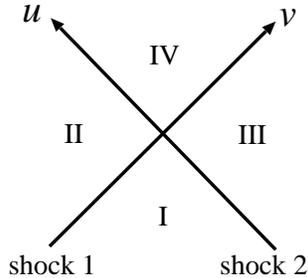}
}
\caption{Schematic picture of the spacetime of colliding high-energy particles. }
\label{shock}
\end{figure}

A quantitative calculation of (the lower bounds on)
$\sigma_{\rm BH}$ in the framework of general relativity
was first done in \cite{EG02} in the four-dimensional case and  was extended
to higher-dimensional cases \cite{YN03} using
a system of colliding Aichelburg-Sexl particles \cite{AS71},
 obtained by boosting the Schwarzschild black hole  to the
speed of light with fixed energy $p$.
A schematic picture of the spacetime is shown in Fig. \ref{shock}.
The gravitational field of each
incoming particle is infinitely Lorentz-contracted and forms a
shock wave. Except at the shock waves, the spacetime is flat
before the collision (i.e., regions I, II, and III). After the
collision, the two shocks nonlinearly interact with each other
and the spacetime within the future lightcone of the collision
(i.e., region IV) becomes highly curved. No one has succeeded
in deriving the metric in region IV even numerically.
However it is possible to investigate the apparent horizon (AH) on the slice $u\le 0=v$
and $v\le 0=u$ and calculate the cross section for AH formation $\sigma_{\rm AH}$
\cite{EG02,YN03}.  $\sigma_{\rm AH}$ provides the lower bound on $\sigma_{\rm BH}$
because AH formation is a sufficient condition for  black hole
formation. Recently one of us and Rychkov \cite{YR05} obtained a somewhat larger
value of $\sigma_{\rm AH}$ by studying the AH on the slice $u\ge 0=v$
and $v\ge 0=u$. The result was $\sigma_{\rm AH}\simeq 3\pi [r_h(2p)]^2$,
e.g., for $D=10$, where $D$ is the total number of spacetime dimensions.

The Aichelburg-Sexl metric describes a
high-energy particle with no charge or spin.
However these quantities  should affect
the formation of black holes because the
gravitational field of each particle is determined by its energy-momentum
tensor. In this paper we consider the effects of electric charge.
Although several interesting discussions
about phenomena associated with charged black hole formation have appeared
in the literature \cite{GT02, CBC05,HKB05},
there has never been a study of black hole formation
resulting from the collision of ultrarelativistic charges.

It is difficult to construct a model of colliding charges that takes account
of the confinement of the electromagnetic field on the brane.
As a first step, we ignore this effect and use
higher-dimensional Einstein-Maxwell theory in order to understand
the features that do not depend on these details.
Furthermore, the Maxwell field would also be higher dimensional
in the neighborhood of the particle if the brane is relatively thick.
We also ignore the brane tension and the structure
of the extra dimensions.

Our approach is as follows.
We model the ultrarelativistic charges using  the metric
obtained by boosting the Reissner-Nortstr\"om black hole
and taking the lightlike limit.
This metric was originally derived by Loust\'o and S\'anchez~\cite{LS90}
and was recently rederived in \cite{Ortaggio06}.
By combining two charges, we set up the head-on collision of
ultrarelativistic charges (whose structure
is similar to Fig.~\ref{shock}) and analyze AH formation
on the slice $u\ge 0 =v$ and $v\ge 0=u$.

Our results indicate that  charge has a significant effect,
typically preventing  black hole formation.
We discuss the implications for  black hole production
in accelerators by choosing parameters appropriate to the LHC.
Taking the boosted Reissner-Nordstr\"om black hole
as a reasonable descriptor of ultrarelativistic charged objects,
charge effects will significantly decrease the rate of black hole formation at accelerators.
However simple order-of-magnitude estimates also show that quantum effects of the
electromagnetic field could play an important role in such situations.
Whether or not they could counteract the effects we obtain remains a subject for
future study.

Our paper is organized as follows.
In the next section, we introduce the boosted Reissner-Nordstr\"om black hole
and set up the system of two ultrarelativistic charges.
In Sec. III, we derive the AH equation and the boundary conditions.
The analytic solution for the AH equation is also represented.
In Sec. IV, we provide the results for the condition for the AH formation
in the system of two ultrarelativistic charges.
Sec. V is devoted to the discussion about the implication of
our results for the black hole production in accelerators.
We conclude with a discussion of effects that we have neglected in this study.


\section{The spacetime of ultrarelativistic charges}

In this section, we study the metric of an ultrarelativistic charge
that is obtained by boosting the Reissner-Nordstr\"om black hole
and introduce the system of two ultrarelativistic charges.

\subsection{Metric of an ultrarelativistic charge}

We begin by reviewing the
ultrarelativistic boost of the Reissner-Nortstr\"om  spacetime metric
in $D$ dimensions originally studied in \cite{LS90}.
The metric of the Reissner-Nordstr\"om spacetime \cite{MP86} is given as
\begin{equation}
ds^2=-g(R)dT^2+g(R)^{-1}dR^2+R^2d\Omega_{D-2}^2,
\end{equation}
\begin{equation}
g(R)=1-\frac{2M}{R^{D-3}}+\frac{Q^2}{R^{2(D-3)}},
\end{equation}
where $Q$ and $M$ are related to charge $q$ and mass $m$ as follows:
\begin{equation}
Q^2=\frac{8\pi G_Dq^2}{(D-2)(D-3)},
\end{equation}
\begin{equation}
M=\frac{8\pi G_Dm}{(D-2)\Omega_{D-2}}.
\end{equation}
Here, $G_D$ is the gravitational constant and $\Omega_{D-2}$ is the
$(D-2)$-dimensional area of a unit sphere.
The electromagnetic field strength tensor $\mathcal{F}_{\mu\nu}$ is given as
\begin{equation}
\mathcal{F}=\frac12E_0 dT\wedge dR,~~E_0=\frac{q}{R^{D-2}}.
\end{equation}

Introducing the isotropic coordinates
$(\bar{T}, \bar{Z}, \bar{r},\bar{\phi}_1,...,\bar{\phi}_{D-3})$,
the metric becomes
\begin{multline}
ds^2=-\frac{\left[\bar{R}^{2(D-3)}-(M^2-Q^2)/4\right]^2}
{\left[\bar{R}^{2(D-3)}+M\bar{R}^{D-3}+(M^2-Q^2)/4\right]^2}d\bar{T}^2
\\
+\left(1+\frac{M}{\bar{R}^{D-3}}+\frac{M^2-Q^2}{4\bar{R}^{2(D-3)}}\right)^{2/(D-3)}\left(d\bar{Z}^2+d\bar{r}^2+\bar{r}^2d\bar{\Omega}_{D-3}^2\right),
\end{multline}
where $\bar{R}=\sqrt{\bar{Z}^2+\bar{r}^2}$
and $d\bar{\Omega}_{D-3}^2$ is the metric of a $(D-3)$-dimensional
unit sphere spanned by $\bar{\phi}_i$.
We apply a boost in the $\bar{Z}$ direction
\begin{align}
\bar{T}&=\gamma(\bar{t}-v\bar{z}), \\
\bar{Z}&=\gamma(-v\bar{t}+\bar{z}),
\end{align}
where $\gamma$ is the Lorentz factor $\gamma=1/\sqrt{1-v^2}$.
We fix both the energy
\begin{equation}
p=m\gamma,
\label{def-p}
\end{equation}
and the following quantity
\begin{equation}
p_e^2=q^2\gamma,
\label{def-pe2}
\end{equation}
and take the lightlike limit $\gamma\to\infty$.
This yields a finite result that
is the charged version of the Aichelburg-Sexl metric \cite{LS90}:
\begin{equation}
ds^2=
-d\bar{u}d\bar{v}+d\bar{r}^2+\bar{r}^2d\bar{\Omega}_{D-3}
+\Phi({\bar{r}})\delta(\bar{u})d\bar{u}^2,
\label{discontinuous}
\end{equation}
\begin{equation}
\Phi({\bar{r}})=
\begin{cases}
\displaystyle -8G_4p\ln \bar{r}-\frac{2a}{\bar{r}}, & (D=4), \\
\displaystyle \frac{16\pi G_Dp}{(D-4)\Omega_{D-3}\bar{r}^{D-4}}
-\frac{2a}{(2D-7)\bar{r}^{2D-7}}, & (D\ge 5),
\end{cases}
\label{potential}
\end{equation}
where
\begin{equation}
a=\frac{2\pi (4\pi G_Dp_e^2)}{(D-3)} \frac{(2D-5)!!}{(2D-4)!!}%
~~(D\ge 4)
\label{avalue}
\end{equation}
and our normalizations of $p$ and $p_e$ differ from those of Ref. \cite{Ortaggio06}.
The metric (\ref{discontinuous}) reduces to the Aichelburg-Sexl metric
in the limit $p_e\to 0$.
The stress-energy tensor has the form
$T_{\mu\nu}=T^{(0)}_{\mu\nu}+T^{\rm (em)}_{\rm \mu\nu}$,
where $T^{(0)}_{\mu\nu}$ and $T^{\rm (em)}_{\rm \mu\nu}$
are proportional to $p\delta(\bar{u})\delta^{D-2}(\bar{r})$ and
$p_e^2\delta(\bar{u})/\bar{r}^{2D-5}$, respectively.
Note that although the value of $q^2$ goes to zero in the infinite  boost limit,
the electromagnetic energy-momentum tensor $T^{\rm (em)}_{\mu\nu}$
has a nonzero distributional value.
For convenience we adopt the quantity
\begin{equation}
r_0=\left(\frac{8\pi G_Dp}{\Omega_{D-3}}\right)^{1/(D-3)}
\label{length-unit}
\end{equation}
as the unit of the length in the following (i.e., $r_0=1$).

{We pause here to comment on the validity and limitations of
the metric \eqref{discontinuous}. First note that when the
the Reissner-Nordstr\"om metric is boosted
in a usual way,  the rest mass $m$ and charge $q$ are fixed
and boosted to a finite value of $\gamma$.
In this case, the leading order contributions of the
mass and the charge to the metric are $O(\gamma m)$
and $O(\gamma q^2)$, respectively.
Setting $p=\gamma m$ and $p_e^2=\gamma q^2$, the boosted metric
approximately coincides with the metric \eqref{discontinuous}--\eqref{avalue},
because the difference is subleading order
$O(m)=O(p/\gamma)$ and $O(q^2)=O(p_e^2/\gamma)$.
Furthermore, in the ultrarelativistic
limit both of these terms diverge unless we take both
$m$ and $q$ to vanish in this limit \cite{LS90}. Since we expect that there will
be energy-momentum due both to mass and to the electromagnetic field of the particle,
it is reasonable to put $p=\gamma m$ and $p_e^2=\gamma q^2$.
Indeed, as noted above the electromagnetic contribution to the
stress-energy of the charged particle has a nonzero (distributional) contribution \cite{LS90}.
Consequently we regard the charged version of the Aichelberg-Sexl metric \eqref{discontinuous}
as a good approximation to an ultrarelativistic massive charged body with finite $\gamma$.
}

Second, there is a restriction on the reliability of the metric
\eqref{discontinuous} that comes from the charge's electrostatic
energy\footnote{
 We thank an anonymous referee for this point.}.
Since the electrostatic energy of a point charge diverges, there
is some radius $r_c$ at which the outside electrostatic energy is
equal to the rest mass of the point charge\footnote{For an
electron in four dimensions, $2r_c\simeq 2.8\text{fm}$ is called
the classical electron radius \cite{Jackson}.}:
\begin{equation}
r_c= \left[\frac{\Omega_{D-2} q^2}{2(D-3)m}\right]^{1/(D-3)}.
\label{classical-radius}
\end{equation}
Because the classical electromagnetism has the contradiction
inside of $r_c$, quantum electrodynamic (QED) effects
become important there. Hence the necessary condition
for the reliability of the metric
\eqref{discontinuous}--\eqref{avalue} is
$\bar{r}\gtrsim r_c$.
Since our analysis of the formation of an AH will be done
around $\bar{r}=r_0/2^{1/(D-3)}$, the necessary condition
for reliability of our results is $r_c^{D-3}\lesssim r_0^{D-3}/2$.
Using Eqs.~\eqref{def-p}, \eqref{def-pe2}, \eqref{avalue} and \eqref{length-unit},
this condition is rewritten as
\begin{equation}
\frac{a}{r_0^{2(D-3)}}
\lesssim \frac{\pi \Omega_{D-3}(2D-5)!!}{\Omega_{D-2}(2D-4)!!}.
\label{model-reliability}
\end{equation}
The value of the right hand side ranges from $0.58$ to $0.68$ for $4\le D\le 11$.

\subsection{Geodesic coordinates}

The delta function in Eq.~\eqref{discontinuous} indicates that
the $(\bar{u},\bar{v},\bar{r})$ coordinate is discontinuous at $\bar{u}=0$.
Seeking new coordinates that are continuous and smooth across the shock,
we introduce  $(u,v,r,\phi_i)$
by the coordinate transformation
\begin{align}
\bar{u}&=u,\\
\bar{v}&=v+F(u,r),\\
\bar{r}&=G(u,r),\\
\bar{\phi}_i&=\phi_i,
\end{align}
where $F(u,r)=0$ and $G(u,r)=r$ for $u<0$.
In the new coordinate system we require that $v,r,\phi_i=\text{const.}$
is a null geodesic with $u$  its affine parameter.
By directly calculating the geodesic equation,
we find that the requirement is satisfied if and only if
\begin{equation}
F_{,u}=G_{,u}^2+\Phi(G)\delta(u),
\label{fu}
\end{equation}
\begin{equation}
F_{,r}=2G_{,u}G_{,r}.
\label{fr}
\end{equation}
are satisfied.
The solution for $F$ and $G$ is
\begin{equation}
F(u,r)=\theta(u)\left[\Phi(r)+\frac{u}{4}\left(\Phi^\prime(r)\right)^2\right]
\end{equation}
\begin{equation}
G(u,r)=r+\frac{u\theta(u)}{2}\Phi^\prime (r),
\end{equation}
where $\theta(u)$ is the unit step function.
The metric in the coordinate $(u,v,r,\phi_i)$ becomes
\begin{equation}
ds^2=-dudv+G_{,r}^2dr^2+G^2d\Omega_{D-3}^2,
\end{equation}
where $G$ and $G_{,r}$ are explicitly given by
\begin{equation}
G=r+\frac{u\theta(u)}{r^{D-3}}\left(1-\frac{a}{r^{D-3}}\right),
\label{funcG}
\end{equation}
\begin{equation}
G_{,r}=1+(D-3)\frac{u\theta(u)}{r^{D-2}}\left(1-\frac{2a}{r^{D-3}}\right).
\end{equation}

\begin{figure}[tb]
 \centering
 \includegraphics[width=0.4\textwidth]{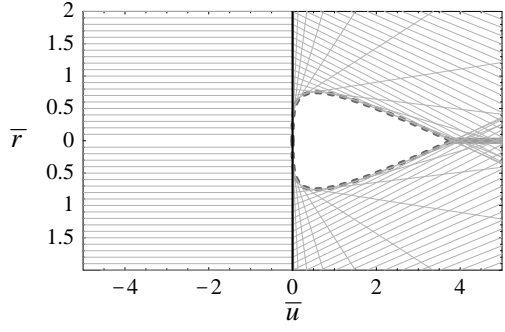}
 \caption{The trajectories of null rays $v, r,\phi_i={\rm const.}$ in the coordinate
 $(\bar{u}, \bar{r})$ in the $D=4$ case. The value of $a$ is $0.75$.
 $\bar{\phi}_i$ and $\bar{v}$ are suppressed. Because the gravitational field
 of the shock is repulsive around the center, the null rays exhibit a crossing
 singularity (the gray dashed line).
 For large $r$, the gravitational field is attractive and the light rays
 will cross at a focusing singularity (the thick gray line). }
 \label{nullray}
\end{figure}

In the coordinate $(u,v,r,\phi_i)$,
the metric coefficients $G^2$ and $G_{,r}^2$ are continuous across the shock.
On the other hand, two coordinate singularities appear in the region $u > 0$.
The first one is
\begin{equation}
u=\frac{-r^{2D-5}}{(D-3)(r^{D-3}-2a)}
\end{equation}
at which $G_{,r}=0$ and the other is
\begin{equation}
u=\frac{r^{2D-5}}{r^{D-3}-a}
\end{equation}
at which $G=0$.
The two singularities cross each other at
$r=r_{\rm c}\equiv\left[(2D-5)a/(D-2)\right]^{1/(D-3)}$.
The light ray $v,r,\phi_i=\text{const}.$ with $r<r_{\rm c}$ will reach the
first singularity $G_{,r}=0$ and the one with $r>r_{\rm c}$
will plunge into the second singularity $G=0$.

 To understand the physical meaning of this it is useful to revert to the
coordinates $(\bar{u}, \bar{v}, \bar{r}, \bar{\phi}_i)$.
Figure \ref{nullray} shows the trajectories of null rays $v,r,\phi_i=\text{const.}$
in the $(\bar{u},\bar{r})$-plane.
In the neighborhood of $r=0$, the gravitational field
of the shock is strongly repulsive and the light rays expand.
Because of this effect, the two neighboring light rays
cross each other.  Since the crossing point corresponds to the point $G_{,r}=0$,
we call it the crossing singularity.
It does not appear in the case of a neutral particle.
For sufficiently large $r$, the gravitational field is attractive and
light rays with the same value of $r$ will be focused on the axis.
This is the point $G=0$ and we call it the focusing singularity.
In the coordinate system $(u,v,r,\phi_i)$, we can only consider the
region prior to the two singularities.

\subsection{The spacetime with two high-energy charges}

Since we have obtained  smooth coordinates for an ultrarelativistic charge,
we can set up a system of two ultrarelativistic charges
as follows.
We assume   without loss of generality that the two particles have the same energy $p$
and different charge parameters $p_e^{(1)}$ and $p_e^{(2)}$.   Using \eqref{avalue}
this implies the two particles have different values of $a$ denoted by $a_1$ and $a_2$.
Because there is no interaction between two particles before the
collision, we simply combine the metric of each particle
in order to obtain the metric of the region outside the
future light cone of the shock collision:
\begin{equation}
ds^2=\left\{
\begin{array}{ll}
-dudv+dr^2+r^2d\Omega_{D-3}^2, & (u\le 0,v\le 0),\\
-dudv+\left(G_{,r}^{(1)}(u,r)\right)^2dr^2+
\left(G^{(1)}(u,r)\right)^2d\Omega_{D-3}^2, & (u\ge 0, v\le 0),\\
-dudv+
\left(G_{,r}^{(2)}(v,r)\right)^2dr^2+
\left(G^{(2)}(v,r)\right)^2d\Omega_{D-3}^2,& (u\le 0, v\ge 0),
\end{array}
\right.
\end{equation}
where $G^{(1)}$ and $G^{(2)}$ are the functions obtained by
substituting $a=a_1$ and $a_2$ for Eq.~\eqref{funcG}, respectively.
Nonlinearities in the field equations obstruct us from obtaining the metric in the region
$u>0, v>0$ .

\section{Finding apparent horizons}

\begin{figure}[tb]
 \centering
 \includegraphics[width=0.5\textwidth]{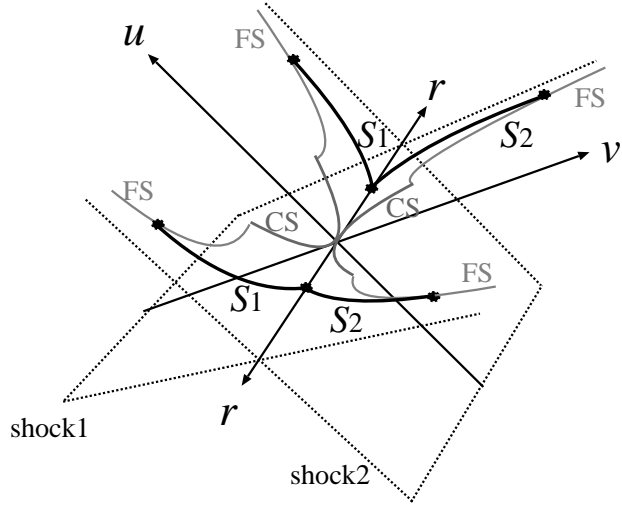}
 \caption{The schematic shape of the AH ($S_1$ and $S_2$ shown by
 thick black lines) on the slice $u>0=v$ and $v>0=u$.
 The crossing singularities (CS) and the focusing
 singularities (FS) are shown by the dark and light gray lines, respectively. $S_1$ and $S_2$
 cross $u=v=0$ at $r=r_{\rm min}$ and the focusing singularities. }
 \label{AH-schematic}
\end{figure}

In this section, we study the AH on the slice $u>0=v$ and $v>0=u$.
Figure~\ref{AH-schematic} shows the schematic shape of the AH
in the slice. Because the system is axi-symmetric,
the location of the AH surface in each side is given by a function of $r$.
We assume that the AH is given by the union of two surfaces $S_1$ and $S_2$ where
\begin{equation}
S_1:
u=h^{(1)}(r) ~~~(r_{\rm min}\le r\le r_{\rm max}^{(1)})~~~\text{on}~~~u>0=v,
\end{equation}
\begin{equation}
S_2:
v=h^{(2)}(r) ~~~(r_{\rm min}\le r\le r_{\rm max}^{(2)})~~~\text{on}~~~v>0=u,
\end{equation}
where $h^{(1)}$ and $h^{(2)}$ are monotonically increasing functions of $r$.
Continuity of the metric at the AH yields the constraint
\begin{equation}
h^{(1)}(r_{\rm min})=h^{(2)}(r_{\rm min})=0
\end{equation}
so that $S_1$ and $S_2$
coincide with each other at $u=v=0$.
At $r=r_{\rm max}^{(n)}$ (where $n=1$ or $2$), we require that $h^{(n)}(r)$
cross the coordinate singularity, i.e.,
\begin{equation}
G^{(1)}(h^{(1)}(r_{\rm max}^{(1)}), r_{\rm max}^{(1)})
=G^{(2)}(h^{(2)}(r_{\rm max}^{(2)}), r_{\rm max}^{(2)})=0.
\end{equation}
Because proper circumference at the focusing singularity
is zero, the surface becomes a closed surface by the above requirements.

\subsection{ AH equation}

Next we derive the equation for $h^{(1)}(r)$ and $h^{(2)}(r)$. Because
the equation for $h^{(2)}(r)$ is obtained by just changing the index
of the equation for $h^{(1)}$, we only have to consider $S_1$.
We put $h(r)=h^{(1)}(r)$ and $G(u,r)=G^{(1)}(u,r)$.

The AH equation is derived by calculating the expansion $\theta_+$
of the null geodesic congruence from the surface and
demanding it to vanish. In order to find the tangent vector $k^\mu$ of the
congruence, we consider the lightcone at each point
on the surface and adopt the outermost one. It is given by
\begin{eqnarray}
k^u&=&\frac{\left(h_{,r}(r)\right)^2}{2\left(G_{,r}(h(r),r)\right)^2}.
\label{ku}
\\
k^v&=&2,\\
k^r&=&\frac{h_{,r}(r)}{\left(G_{,r}(h(r),r)\right)^2},
\label{kr}
\\
k^{\phi_i}&=&0.
\end{eqnarray}
By calculating the evolution of the area along the congruence,
we find that the expansion $\theta_+$ is
\begin{equation}
\theta_+=\partial_rk^r+(D-3)\frac{G_{,u}k^u+G_{,r}k^r}{G}
+\frac{G_{,ru}k^u+G_{,rr}k^r}{G_{,r}}.
\end{equation}
Substituting Eqs. \eqref{ku} and \eqref{kr} and imposing $\theta_+=0$, we find
\begin{equation}
h_{,rr}+h_{,r}
\left[
(D-3)\frac{(1/2)G_{,u}h_{,r}+G_{,r}}{G}
-\frac{(3/2)G_{,ru}h_{,r}+G_{,rr}}{G_{,r}}
\right]=0.
\label{AH-equation}
\end{equation}
This is the AH equation.

\subsection{Boundary conditions}

The continuity of the tangent vector $k^\mu$ of the congruence
should be imposed at $r=r_{\rm max}$ and $r=r_{\rm min}$.
Otherwise, the surface would have a delta function expansion
and would not satisfy the AH condition \eqref{AH-equation}.

At $r=r_{\rm max}$, we return to the coordinates
$(\bar{u}, \bar{v}, \bar{r}, \bar{\phi}_i)$ and impose $k^{\bar{r}}=0$.
This is equivalent to
\begin{equation}
h_{,r}(r_{\rm max})=
\frac{-2G_{,r}(h(r_{\rm max}), r_{\rm max})}{G_{,u}(h(r_{\rm max}), r_{\rm max})}.
\end{equation}
From the AH Eq. \eqref{AH-equation}, we see that
this is equivalent to the regularity condition at the focusing singularity $G=0$.
Hence, if we find a regular solution of Eq.~\eqref{AH-equation}
that crosses the focusing singularity, it will automatically satisfy the
boundary condition at $r=r_{\rm max}$.

In order to find the boundary condition at $r=r_{\rm min}$,
we consider both $S_1$ and $S_2$.
The tangent vectors $k_1^\mu$ and $k_2^\mu$ of the congruence of
surfaces $S_1$ and $S_2$ are
given by
\begin{equation}
k_1^v=2, ~~k_1^u=\left(h_{,r}^{(1)}\right)^2/2,
\end{equation}
\begin{equation}
k_2^v=\left(h_{,r}^{(2)}\right)^2/2, ~~k_2^u=2,
\end{equation}
respectively, at $r=r_{\rm min}$. Because $k_1^\mu$
and $k_2^\mu$  point in the  same direction,
$k_1^uk_2^v=k_1^vk_2^u$ holds.
This is equivalent to
\begin{equation}
h_{,r}^{(1)}(r_{\rm min})h_{,r}^{(2)}(r_{\rm min})=4.
\label{innerBC}
\end{equation}
which is the boundary condition that must be imposed  at $r=r_{\rm min}$.
Note that both $h_{,r}^{(1)}(r_{\rm min})$ and $h_{,r}^{(2)}(r_{\rm min})$
are positive.

\subsection{Solutions}

The AH Eq.~\eqref{AH-equation} can be solved exactly, yielding
a one parameter
family of regular solutions given by
\begin{equation}
h(r)=
\frac{2r^{2}}{\left( 1-{a}/{r}\right) ^{2}}\left[ \ln \left(
\frac{r}{r_{\rm min}}\right) +{a}\left(\frac{1}{r}-\frac{1}{r_{\rm min}}\right)\right],
\end{equation}
for $D=4$ and
\begin{equation}
h(r)=\frac{2}{(D-4)}\frac{r^{D-2}}{\left( 1-{a}/{r^{D-3}}\right) ^{2}}
\left[ \left(1-\frac{D-4}{2D-7}\frac{a}{r_{\rm min}^{D-3}}\right)
\left( \frac{r}{r_{\rm min}}\right) ^{D-4}
-1+\frac{D-4}{2D-7}\frac{a}{
r^{D-3}}\right],
\end{equation}
for $D\ge 5$.
These solutions satisfy the boundary condition at $r=r_{\rm max}$
and $h(r_{\rm min})=0$. The quantity
$h_{,r}(r_{\rm min})$ becomes
\begin{equation}
h_{,r}(r_{\rm min})=\frac{2x^2}{x-a},
\end{equation}
where
\begin{equation}
x\equiv r_{\rm min}^{D-3}.
\end{equation}
Then, the boundary condition \eqref{innerBC} becomes
\begin{equation}
x^4=(x-a_1)(x-a_2).
\label{equation-rmin}
\end{equation}
This equation determines the value of $r_{\rm min}$; indeed
the AH exists if and only if there is a solution to Eq.~\eqref{equation-rmin}.
Note that $x$ must be larger than $a_1$ and $a_2$ because
$h_{,r}(r_{\rm min})$ is positive.

\section{Results}

In the study in the previous section, the problem of finding the AH was
reduced to solving the quartic equation \eqref{equation-rmin}.
Now we study the condition for the AH existence.

\subsection{Collision of charges with the same $a$}

 As a concrete example, we first consider the situation where both
charges are the same, i.e., $a_1=a_2=a$.
In this case, $x$ is solved as
\begin{equation}
x=\frac{1\pm\sqrt{1-4a}}{2}.
\end{equation}
The AH exists only when $a\le 1/4$.

\begin{figure}[tb]
 \centering
 \includegraphics[width=0.3\textwidth]{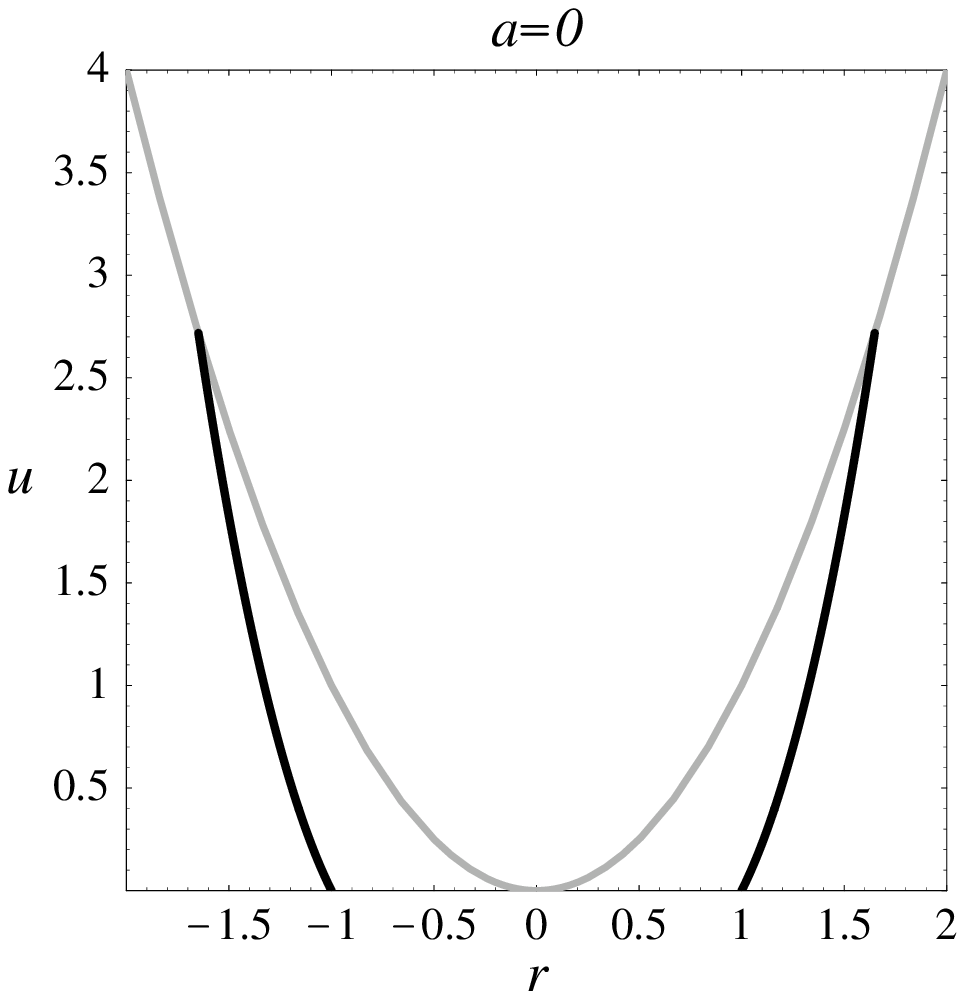}
 \includegraphics[width=0.3\textwidth]{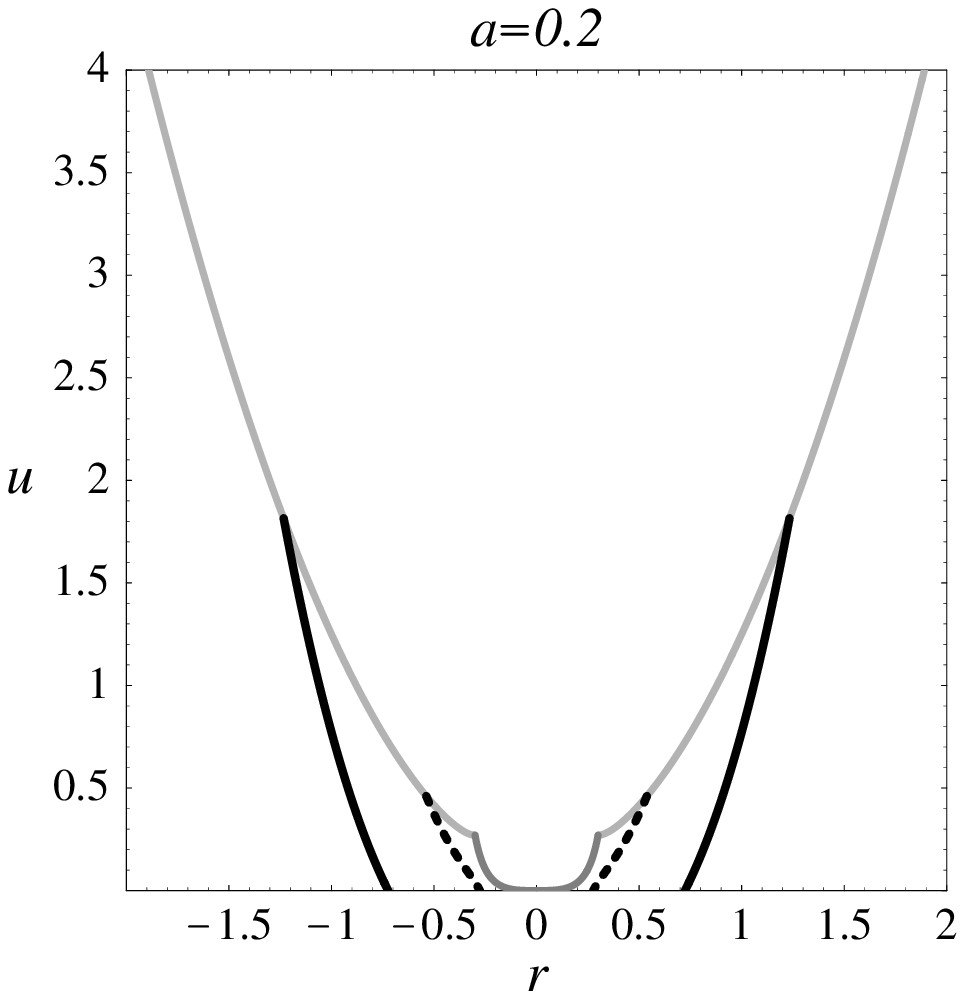}
 \includegraphics[width=0.3\textwidth]{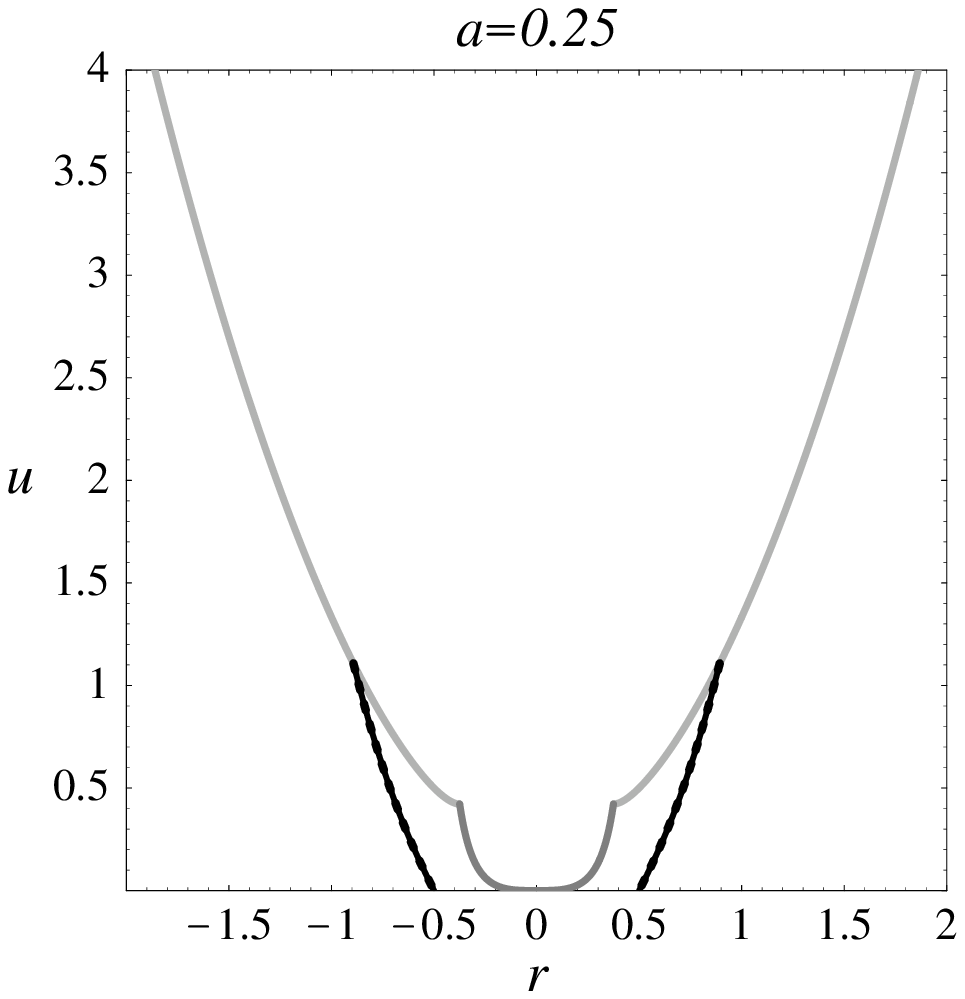}
 \caption{The shape of the AH (the black lines) that is formed
 in the collision of two charges with the same $a$ for $D=4$.
 The values of $a$ are $0,0.2,0.25$. The inner boundary
 of the trapped region is also shown by dashed lines.
 The gray lines indicate the coordinate singularities.
 For $a>0.25$, the trapped region disappears.}
 \label{AHshape-qq}
\end{figure}

In Fig. \ref{AHshape-qq}, we show the examples of the solutions in the $D=4$ case
for $a=0,1/5,1/4$.
For $a>0$, there are two solutions that correspond to the inner
and outer boundaries of the trapped region.
If we increase $a$, the trapped region shrinks and
the two solutions become degenerate at $a=1/4$.
There is no AH for $a>1/4$.

\subsection{Collision of a charged and a neutral particle}

\begin{figure}[tb]
 \centering
 \includegraphics[width=0.25\textwidth]{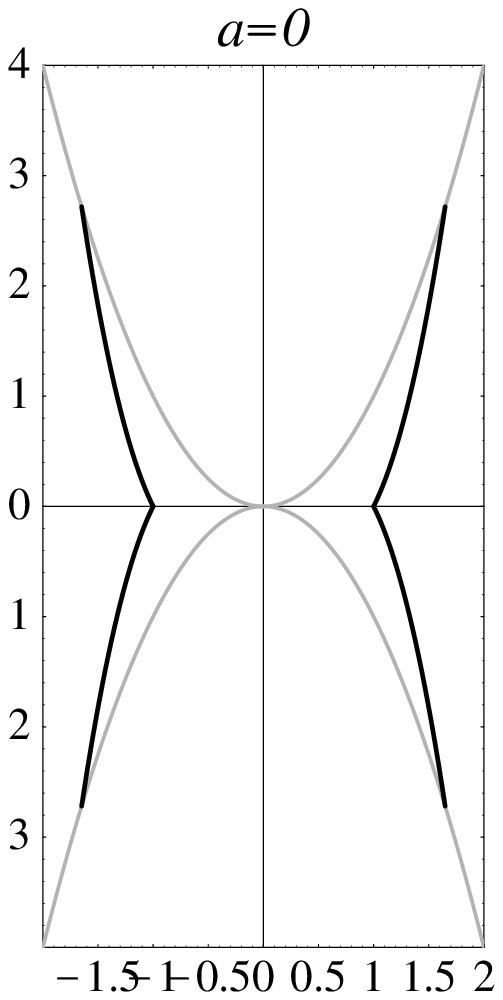}
 \includegraphics[width=0.25\textwidth]{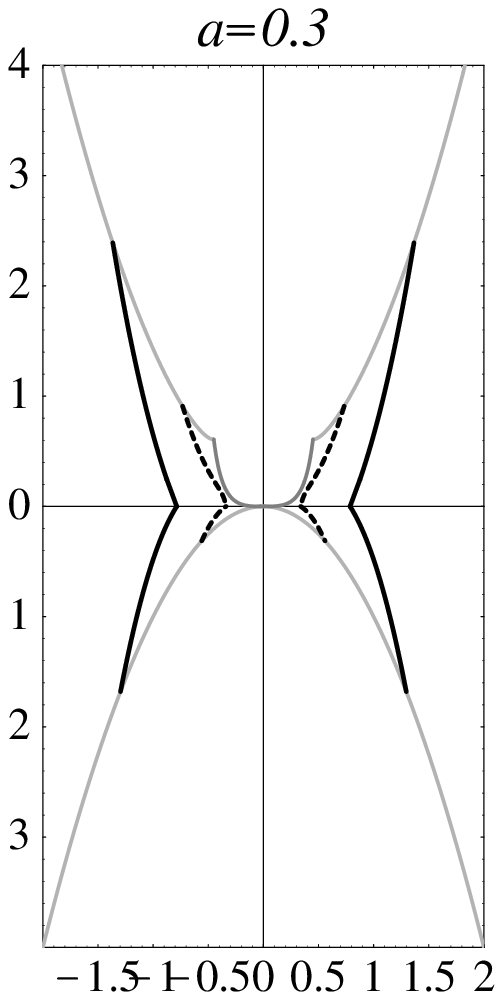}
 \includegraphics[width=0.25\textwidth]{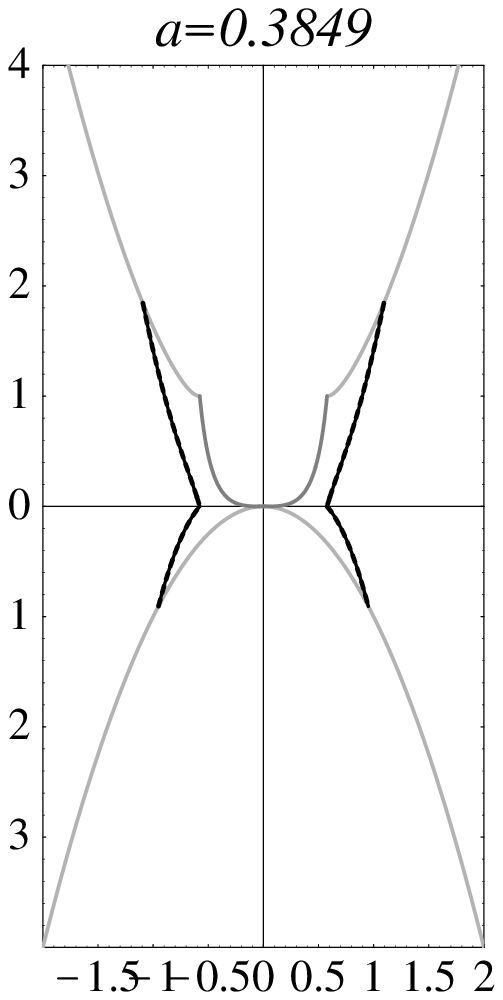}
\caption{The shape of the AH (the black line) that is formed
 in the collision of a charge with parameter $a$ and a neutral particle in the $D=4$ case.
 $S_1$ is shown in the upper side and $S_2$ is shown in the lower side.
 The values of $a$ are $0,0.3,0.3849$. }
\label{AHshape-q0}
\end{figure}

Next, we consider the case where one particle has $a_1=a$
and the other is neutral, i.e., $a_2=0$.
In this case, Eq.~\eqref{equation-rmin} becomes
\begin{equation}
x^3-x+a=0.
\end{equation}
Solutions exist only for $a\le 2/(3\sqrt{3})$.

Figure \ref{AHshape-q0}
shows the shape of the AH in the $D=4$ case.
Similarly to the same charge case,
two solutions appear and they degenerate at $a=\frac{2}{3\sqrt{3}}\simeq 0.3849$.
For $a>\frac{2}{3\sqrt{3}}$, there is no AH.

\subsection{General charge parameters}

\begin{figure}[tb]
\centering
{
\includegraphics[width=0.4\textwidth]{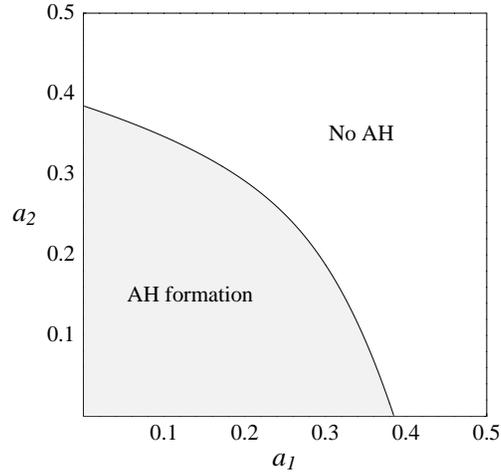}
}
\caption{The region of the AH formation in the $(a_1, a_2)$-plane.}
\label{region}
\end{figure}

Now we consider the condition for  AH formation
for general $a_1$ and $a_2$.
In general, Eq. \eqref{equation-rmin} has four solutions.
Under the condition $a_1>0$ and $a_2>0$, there is always one negative
solution and one positive solution smaller than $\text{min}[a_1,a_2]$.
However, these two solutions do not correspond to an AH
because $x$ must be greater than $a_1$ and $a_2$.

We therefore investigate the existence of the other two solutions.
Setting
\begin{equation}
f(x)=x^4-(x-a_1)(x-a_2),
\end{equation}
we find that an AH exists if and only if
the local minimum of $f(x)$ in the $x>0$ region is less than or
equal to zero.
The location of the local minimum is given by the following
equation:
\begin{equation}
\frac{df}{dx}=4x^3-2x+(a_1+a_2)=0,
\end{equation}
whose positive solution is
\begin{equation}
x=\frac13\left[-(a_1+a_2)+\sqrt{(a_1+a_2)^2-8/27}\right]^{-1/3}
+\frac12\left[-(a_1+a_2)+\sqrt{(a_1+a_2)^2-8/27}\right]^{1/3}.
\end{equation}
Substituting this value into $f(x)$ and drawing the contour for $f(x)=0$,
we find the region for the AH formation on the $(a_1,a_2)$-plane,
shown in Fig.~\ref{region}. Both $a_1$ and $a_2$ must be
sufficiently small for  AH formation.

\subsection{Physical interpretation}

Since $a_1$ and $a_2$ are proportional to $(p_e^{(1)})^2$
and $(p_e^{(2)})^2$, the condition derived above does not
depend on the sign of the charge of either particle. 
This is because the
gravitational field due to each charge is
generated by an electromagnetic energy-momentum tensor $T^{\rm (em)}_{\mu\nu}$
that depends on the squared charge.
As pointed out in Sec. II, the gravitational field induced by
$T^{\rm (em)}_{\mu\nu}$ of the incoming particles is repulsive, and
its effect becomes dominant around the center.
As the value of $a$ increases, the repulsive region becomes
large, preventing formation of the AH .

 This effect is reminiscent of that
found in the original Reissner-Nordstr\"om black hole. If we
increase the charge $|q|$, the inner and outer horizons become
closer, coalescing at $|q|=\frac{m}{\Omega_{D-2}}\sqrt{\frac{8\pi
(D-3)G_D}{(D-2)}}$. This example also makes clear that the
gravitational field generated by a Coulomb field is repulsive
regardless of the sign of the charge and tends to obstruct black
hole formation. However the analogy with the Reissner-Nordstr\"om
black hole has its limitations. One might naturally expect that a
black hole forms from the collision of two charges $q_1$ and $q_2$
when and only when the total charge $|q_1+q_2|$ of the system is
less than $\frac{(2p)}{\Omega_{D-2}}\sqrt{\frac{8\pi
(D-3)G_D}{(D-2)}}$ where $2p$ is the total system energy. However
our analysis shows that an AH does not necessarily form even in
the case $|q_1+q_2|\ll \frac{(2p)}{\Omega_{D-2}}\sqrt{\frac{8\pi
(D-3)G_D}{(D-2)}}$, because  the repulsive gravitational
effect due to the electric field is enhanced by a factor of
$\gamma$. The critical value of $a$ for AH formation is where
this enhanced repulsive force becomes equivalent to the
self-attractive force due to the energy of the system. We will
return to this point in the next section when we apply our results
to the LHC phenomena.

Note that there remains a possibility that a black hole will form
upon collision even if there is no AH on the slice we have
studied, because  AH formation is only a sufficient condition for
black hole formation.  While a study of the temporal evolution
after collision is beyond the scope of this paper, let us briefly
discuss how the condition for black hole formation is expected to
be  modified from relative to the condition for AH formation.

In the collision between a charged and a neutral particle, the
gravitational interaction  begins after the collision and there is
no electromagnetic interaction between the two particles.
Because the gravitational field of a charged particle remains
repulsive, the condition for the black hole formation is also
given by $a\lesssim 2/(3\sqrt{3})$ up to a factor close to unity.
In the case of a collision between particles with equal charge,
electromagnetic interactions will begin after the collision.
Both the gravitational and electromagnetic forces acting between
the two particles are repulsive; there is no reason to expect that
the electromagnetic interaction enhances black hole formation.
Modifications from the condition for AH formation are again
expected to be small and the condition for the black hole
formation should also be $a\lesssim 1/4$.

In a collision between two particles with charges of opposite sign
these results might change. The electromagnetic force acting
between two particles after the collision is attractive in this
case. The Coulomb fields of the two particles will tend to cancel
each other, suppressing the repulsive force of  the contributions
to the gravitational field due to each charge. If this effect is
significant, the electromagnetic interaction will enhance black
hole formation and it is conceivable that the condition for the
black hole formation might significantly differ from $a<1/4$.
Since we cannot know how effective the electromagnetic interaction
is in enhancing black hole formation without directly computing
the subsequent temporal evolution, we leave the derivation of the
correct condition for future research.

The electromagnetic radiation emitted in the collision
process crucially depends on whether a black hole forms or not. If
a black hole forms, it will shed hair by radiating gravitational
and electromagnetic waves.  An investigation of the ratio of
these two kinds of radiated energy in a slightly different setup
found that the electromagnetic energy is suppressed relative to
the gravitational energy for very high energies \cite{CLY03}. We
expect that similar results hold for our system, because most of
the bremsstrahlung radiation due to the collision will be hidden
inside of the horizon. On the other hand, if a black hole does not
form, the scattering of two particles will occur. In this case,
strong bremsstrahlung radiation is expected because the
electromagnetic interaction is not hidden inside of the horizon.

We now discuss the reliability of our results on the condition for
AH formation. In Sec. IIA, we introduced the radius $r=r_c$ at
which the exterior electrostatic energy becomes equal to the rest
mass. The necessary condition for the reliability of our model is
given by Eq.~\eqref{model-reliability}, which is approximately
$a\lesssim 0.6$. Hence the range of $a$ where the AH is prohibited
is $1/4\le a\lesssim 0.6$ in the case of the collision of two
particles with equal charge. However $r<r_c$ is a sufficient
condition for the importance of QED effects, but is not a
necessary condition. It is possible that QED effects are important
also in a neighborhood of $r\sim r_c$. An exact description of
ultrarelativistic collisions of charged bodies thus entails
inclusion of QED effects. However the discussion above indicates
that we cannot be sure if QED effects suppress or enhance the
repulsive effect we have obtained.

In the next section, we discuss what phenomena will result at the
LHC assuming QED effects are small and thus our results are
applicable.

\section{Discussion}

We discuss here possible implications of our results
in the context of the TeV gravity scenarios
by evaluating the characteristic value of $a$ in future accelerators,
assuming that boosted Reissner-Nordstr\"om black holes
represent the gravitational field of  elementary particles with
electric charge moving at high speed.
We also discuss the reliability of our results
by discussing other possible effects
that are not included in our current analysis.

To simplify the discussion, let us consider the
case of a head-on collision of
two particles with equal charge and mass.
In this case, the condition for the AH formation
in the head-on collision is given by $a/r_0^{2(D-3)}\le 1/4$,
where we have restored the length unit $r_0$. This is equivalent to
\begin{equation}
\frac{p_{e}^{2}}{G_Dp^{2}}\le \frac{2(D-3)}{\Omega_{D-3}^2}
\frac{(2D-4)!!}{(2D-5)!!}.  \label{eq:condition_AH}
\end{equation}
Because both $p_e^2$ and $p$ are proportional  to the Lorentz factor $\gamma$,
the left hand side goes to zero and charge effects are
not significant at high energies.
But is the energy sufficiently high at the LHC?

In order to evaluate the value of $p_e^2$, we
use the original definiton $p_e^2=\gamma q^2$,
where $q^2$ is the squared charge in the higher-dimensional Maxwell theory.
We first establish the relationship between the higher-dimensional charge $q^2$ and the
four-dimensional one $q_4^2$.
For this purpose, we consider the two particles with
the same charge $q$ at rest. The force acting between two particles is given
by
\begin{equation}
F=\frac{q^2}{r^{D-2}}.
\end{equation}
If we assume that the gauge field is confined on the brane, the unique
characteristic length scale is the width of the brane, which should be of
the order of the Planck length $(C_{\mathrm{brane}}/M_p)$,
where $M_p$ is the Planck mass and $C_{\mathrm{brane}}$ is a
dimensionless quantity of order one.
Hence, for sufficiently large $r$, $F$ becomes
\begin{equation}
F\to \frac{q^2}{r^2}\left(\frac{M_p}{C_{\mathrm{brane}}}\right)^{D-4} =\frac{%
q_4^2}{r^2},
\end{equation}
and we find
\begin{equation}
q^2=q_4^2\left(\frac{C_{\mathrm{brane}}}{M_p}\right)^{D-4}.
\label{q-q_4-relation}
\end{equation}
For the characteristic value of $q_4^2$, we adopt
\begin{equation}
q_4^2=C_q^2\alpha,
\end{equation}
where $\alpha$ is the fine structure constant and $C_q$
is the charge in units of the elementary charge $e$.
In the quark case, $C_q$ is $1/3$ or $2/3$.
We also rewrite the gravitational constant $G_D$ in
terms of the Planck mass $M_p$.
Notwithstanding several definitions of the Planck mass
(summarized in \cite{Giddings01}),
we adopt the definition
\begin{equation}
G_D^{-1}=\frac{4\pi}{(2\pi)^{D-4}}M_p^{D-2}.
\label{Planck}
\end{equation}
Then the condition is rewritten as
\begin{equation}
C_q^2\alpha \left(\frac{M_p}{m}\right)\left(\frac{M_p}{p}%
\right)\lesssim \frac{(D-3)}{2\pi\Omega_{D-3}^2}\frac{(2D-4)!!}{(2D-5)!!}
\left(\frac{2\pi}{C_{\rm brane}}\right)^{D-4},
\label{formation-condition}
\end{equation}
where $m$ denotes the rest mass of each incoming particle.

\begin{table}[tb]
\centering
\caption{The values of the right hand side of Eq.~\eqref{formation-condition}
for two cases $C_{\rm brane}=1$ and $2\pi$ . }
\label{righthandside}
\begin{ruledtabular}
\begin{tabular}{c|cccccccc}
$D$ & $4$ & $5$ & $6$ & $7$ & $8$ & $9$ & $10$ & $11$  \\
  \hline
$C_{\rm brane}=1$ & $0.01$ & $0.04$ & $0.2$ & $1$ & $6$ & $40$ & $300$ & $3000$ \\
$C_{\rm brane}=2\pi$ & $0.01$ & $0.006$ & $0.004$ & $0.004$ & $0.004$ & $0.004$ & $0.005$ & $0.008$ \\
  \end{tabular}
  \end{ruledtabular}
\end{table}

The values of the right hand side are summarized in the
cases $C_{\rm brane}=1$ and $2\pi$ in Table \ref{righthandside}.
In the case $C_{\rm brane}=1$, it is less than $1$ for $D\le 7$
and becomes larger as $D$ increases.
In the case $C_{\rm brane}=2\pi$, it is less than $0.01$ for all $5\le D\le 11$.
On the other hand, the natural values for the factors in the left hand side
at the LHC would be $\alpha\simeq1/137$
\footnote{ If we take the running
of the coupling constant into account, a somewhat larger value  $\alpha\simeq1/120$
might be better. Adopting this value, the condition of the black hole formation
will become a bit stricter. },
$(M_p/m)\sim\mathrm{1TeV}/\mathrm{5MeV}=200000$
(for a up or down quark) and $(M_p/p)=1/\mathrm{few}$.
In quark collisions at the LHC,  AH formation will not occur
at the instant of collision,
if the brane is somewhat thick or if the dimensionality $D$ is not too large.

\begin{table}[tb]
\centering
\caption{The values of the right hand side of Eq.~\eqref{extremal-condition}
for two cases: $C_{\rm brane}=1$ and $2\pi$. }
\begin{ruledtabular}
\begin{tabular}{c|cccccccc}
$D$ & $4$ & $5$ & $6$ & $7$ & $8$ & $9$ & $10$ & $11$  \\
  \hline
$C_{\rm brane}=1$ & $0.006$ & $0.02$ & $0.09$ & $0.4$ & $2$ & $20$ & $100$ & $1000$ \\
$C_{\rm brane}=2\pi$ & $0.006$ & $0.003$ & $0.002$ & $0.002$ & $0.002$ & $0.002$ & $0.002$ & $0.003$ \\
  \end{tabular}
  \end{ruledtabular}
\end{table}

Until now charge effects were presumed to be small~\cite{GT02},
because they were expected to be proportional to the fine
structure constant $\alpha\simeq 1/137$. Our analysis indicates
that charge effects can be quite large, because the
electromagnetic energy-momentum tensor $T^{\rm (em)}_{\mu\nu}$ is
proportional to $p_e^2\sim \gamma\alpha$ and the Lorentz factor
$\gamma$ is much larger than $1/\alpha$ for  ultrarelativistic
charges  and so our results are quite nontrivial. Actually
charged black  holes were expected to form in collisions
between ultrarelativistic charges at the LHC \cite{GT02,
CBC05,HKB05}, because a black hole with mass few TeV and
elementary charge $e$ is able to exist, as shown in the following.
The condition for the existence of an horizon in the
Reissner-Nordstr\"om spacetime is $ |q|\le
\frac{m}{\Omega_{D-2}}\sqrt{\frac{8(D-3)\pi G_D}{(D-2)}}. $ Using
the definition of the Planck mass \eqref{Planck} and Eq.
\eqref{q-q_4-relation}, we find that this condition becomes
\begin{equation}
q_4^2\left(\frac{M_p}{m}\right)^2
\le \frac{2(D-3)}{(D-2)\Omega_{D-2}^2}\left(\frac{2\pi}{C_{\rm brane}}\right)^{D-4}.
\label{extremal-condition}
\end{equation}
Setting $q_4^2=\alpha=1/137$ and $m=\text{few}\times M_p$, the
typical value of the left hand side becomes $\sim 0.001$. The
values of the right hand side are given in  Table II. The
inequality \eqref{extremal-condition} is satisfied and there is
indeed a black hole\footnote{Whether a charged black hole can
exist in the ADD scenario was first discussed in \cite{CH02} using
the Reissner-Nordstr\"om metric. There it was stated that an
elementary-charged object a few TeV in mass should be a naked
singularity because it violates the condition admitting the
existence of an horizon. In this discussion, however, the effect
of gauge-field confinement on the brane seems to have been
ignored. }, although it is in a near extremal state in the thick
brane case. So the important point of our results is that the
collision of two charges $q_1$ and $q_2$ with the center-of-mass
energy $2p$ does not necessarily lead to black hole formation even
if there exists a black hole of mass $2p$ and charge $|q_1+q_2|$,
because the charge effect is enhanced by a factor of $\gamma$.

Keep in mind that the condition for the black hole formation is
different from the condition for AH formation. As discussed in
Sec. IVD, however, two conditions become similar in the case of a
collision of particles whose charges have the same sign or a
collision between a charged and a neutral particle.  There is a
possibility that the condition for black hole formation is
significantly modified from the condition for AH formation in the
case of a collision between two equal but opposite-signed charged
particles, because the electromagnetic interaction can enhance
gravitationally attractive effects contributing to black hole
formation. However even if we assume that a black hole forms in
this case, black hole production occurs only when a quark and its
antiquark (or two gluons\footnote{ Although the gluons do not
have  electric charge, they have color charge. Hence we should
note that if color charge has an effect analogous to that which
we have found in this paper, black holes would not be produced
also in gluon collisions.}) collide at the LHC. Then black hole
formation would become a rare process -- just scattering with the
associated bremsstrahlung radiation would occur in most cases. As
a result, the black hole production rate could significantly
decrease relative to previous expectations, and detecting  black
hole signals will be much more difficult than originally expected.
In order to specify more detailed criteria for black hole
formation it will be necessary to study temporal evolution of the
spacetime after collision. Inclusion of brane effects on  gauge
field confinement may also be necessary.

We note also that additional effects such as inclusion of the
spin of incoming particles are also required. If such effects also
weaken the gravitational field of incoming particles in a manner similar
to that of electric charge,
the black hole production rate at the LHC
might further decrease. Inclusion of spin might be carried out via
the interesting ``gyraton'' model~\cite{gyraton}  proposed recently.
This is the spacetime of a source
of finite width propagating at the speed of light.
It has internal angular momentum (spin) and its gravitational
field produces a  frame-dragging effect. Since its gravitational field is also repulsive
around its center we expect similar inhibition of  AH formation
to be observed in the gyraton collisions.

Finally, we revisit the issue of reliability of the charged
particle model that we used. As we  pointed out in Sec. IIA and
IVD, QED effects become important within the radius $r_c$ where
the exterior electrostatic energy is equal to the rest mass
energy. Substituting the relation \eqref{q-q_4-relation} into
Eq.~\eqref{classical-radius}, $r_c$ is given by
 \begin{equation}
r_c\sim C_{\rm brane}^{(D-4)/(D-3)}
\left(\frac{q_4^2}{mM_p^{D-4}}\right)^{1/(D-3)}.
\end{equation}
For a quark, $r_c$ ranges from $10^{-4}$--$10^{-3} \text{fm}$ and
is larger than $r_h(2p)\sim 10^{-4}{\rm fm}$.  Consequently our AH
analysis was carried out in a regime where the quantum
electrodynamic effects may be important. In order to study QED
effects, it will be necessary to derive the gravitational field of
a charge produced by the expectation value of the energy-momentum
tensor $\langle T_{\mu\nu}^{\rm (em)}\rangle$. One realization of
QED effects is the existence of a nonzero trace anomaly $\langle
T_{~~~~\mu}^{{\rm (em)}\mu}\rangle$ \cite{trace-anomaly}. Such QED
effects should affect the condition for the AH formation in
ultrarelativistic charged collisions, an interesting subject for
 future study.
While we cannot definitely conclude that charge inhibits (or even
prevents) black hole formation at the LHC, it is not obvious
whether the QED effects weaken or further strengthen the repulsive
effect we obtained.

To summarize, we studied AH formation in the
head-on collision of the ultrarelativistic charges,
 modeled by boosted Reissner-Nordstr\"om
black holes. We find that charge inhibits the formation of an AH.
Our results suggest
a decrease of the black hole production rate
at the LHC in TeV gravity scenarios, although further studies on the
evolution after the collision and QED effects are required.

\acknowledgments

H.Y. thanks Tetsuya Shiromizu, Antonino Flachi, Hideo Kodama and Masayuki Asakawa
for helpful comments. R.B.M. is grateful for the hospitality of TiTech, where this work was
initiated. The work of H.Y. was partially supported by a Grant for The 21st Century
COE Program (Holistic Research and Education Center for Physics
Self-Organization Systems) at Waseda University. R.B.M. was supported in
part by the Natural Sciences and Engineering Research Council of Canada.

\end{document}